\documentclass[graybox]{svmult}
\usepackage{mathptmx}       % selects Times Roman as basic font
\usepackage{helvet} 

\usepackage{makeidx}         % allows index generation
\usepackage{graphicx}        % standard LaTeX graphics tool
\usepackage{multicol}        % used for the two-column index
\usepackage[bottom]{footmisc}% places footnotes at page bottom
\usepackage{amsmath,amssymb,amsfonts}        % AMS math packages
\usepackage{braket}
\usepackage{cite}
\usepackage{xcolor}
\newcommand{\ttb}{\mathrm{T}\overline{\mathrm{T}}}
\newcommand{\fnc}{\mathbf{F}}
\makeindex             % used for the subject index
\definecolor{darkblue}{rgb}{0.0, 0.27, 0.63}  % Darker blue color definition

\usepackage[colorlinks=true, linkcolor=blue, citecolor=blue, urlcolor=blue]{hyperref}

\title*{A Note on $\ttb$ Deformations and Boundaries}
\author{Nicolò Brizio, Tommaso Morone and Roberto Tateo}
\institute{Nicolò Brizio \at Dipartimento di Fisica, Università di Torino and \\ INFN, Sezione di Torino, Via P. Giuria 1, 10125, Torino, Italy \\ \email{nicolo.brizio@unito.it}
\and Tommaso Morone \at Dipartimento di Fisica, Università di Torino and \\ INFN, Sezione di Torino, Via P. Giuria 1, 10125, Torino, Italy \\ \email{tommaso.morone@unito.it}
\and Roberto Tateo \at Dipartimento di Fisica, Università di Torino and \\ INFN, Sezione di Torino, Via P. Giuria 1, 10125, Torino, Italy \\ \email{roberto.tateo@unito.it}
}
\begin{document}
\maketitle
\abstract{
The irrelevant composite operator $\ttb$, constructed from components of the stress-energy tensor, exhibits unique properties in two-dimensional quantum field theories and represents a distinctive form of integrable deformation. Significant progress has been made in understanding the bulk aspects of the theory, including its interpretation in terms of coordinate transformations and its connection to topological gravity models. However, the behavior of $\ttb$-deformed theories in the presence of boundaries and defects remains largely unexplored. In this note, we review analytical results obtained through various techniques. Specifically, we study the $\ttb$-deformed exact $g$-function within the framework of the Thermodynamic Bethe Ansatz and show that the results coincide with those obtained by solving the corresponding Burgers-type flow equation. Finally, we highlight some potentially significant open problems.
}

\section{Introduction and Overview}

The investigation of thermodynamics in one-dimensional integrable models with factorized scattering originates from the pioneering work of Yang and Yang \cite{Yang:1966ty, Yang:1968rm}. Their formulation, now referred to as the Thermodynamic Bethe Ansatz (TBA), has proven to be remarkably flexible, finding application across a wide range of models \cite{Zamolodchikov:1989cf, Klassen:1989ui, Klassen:1990dx, Klumper:1991jda, Destri:1992qk, Destri:1994bv, Dorey:1996re}, and leading to exact equations for the free energy.

In realistic physical systems, impurities, domain walls, and defects that break translational and rotational symmetries into subgroups are ubiquitous \cite{Andrei:2018die}. Methods from integrability provide robust tools for addressing such complexities. A particularly relevant notion is the $g$-function, which extends the notion of non-integer ground state degeneracy linked to critical boundary conditions, initially introduced by Affleck and Ludwig for the Kondo model \cite{Affleck:1991tk, Affleck:1990iv}. The $g$-function quantifies the contribution of a single boundary to the free energy, and represents a pivotal tool in analyzing renormalization group-type flows in boundary field theories.

Two distinct Hamiltonian representations exist for the partition function on a finite cylinder. In the first formulation, the Hamiltonian governs the time evolution of states within a Hilbert space defined on an interval of length $L$, with boundary conditions $w_1$ and $w_2$ imposed at the endpoints. Alternatively, the time evolution can be understood as corresponding to the height $L$ of the cylinder, involving boundary states $\ket{w_1}$ and $\ket{w_2}$, along with the eigenstates $\left\{\ket{\psi_n}\right\}$ of the Hamiltonian on a circle of radius $R$. The equivalence of these two formulations yields the following fundamental identity:
\begin{equation}\label{channel}
\sum_{n\ge0}e^{-RE_n^\text{strip}(L)}=\sum_{n\ge0}\frac{\braket{w_1|\psi_n}\braket{\psi_n|w_2}}{\braket{\psi_n|\psi_n}}e^{-LE_n^\text{circ}(R)}\,,
\end{equation}
and
\begin{equation}
\text{ln}\,g_{n,\ket{w_1}}=\text{ln}\frac{\braket{w_1|\psi_n}}{\braket{\psi_n|\psi_n}^{1/2}}\,.
\end{equation}
In particular
$\text{ln}\,g_{\ket{w_1}} :=\text{ln}\,g_{0,\ket{w_1}}$
is the so-called ground-state $g$-function.
For integrable models, in the Thermodynamic Bethe Ansatz framework, an exact expression for the $g$-function can be obtained for large families of quantum field theories \cite{LeClair:1995uf, Dorey:2004xk, Dorey:2009vg, Dorey:2005ak, Dorey:2010ub,Caetano:2020dyp}. 

Besides boundary conditions, studying renormalization group flows from non-trivial fixed points remains a primary area of interest.

Deformations of quantum field theories (QFTs) are classified based on the nature of the perturbing operators: relevant, marginal, or irrelevant \cite{RevModPhys.47.773}. While relevant and marginal deformations have been extensively studied, irrelevant deformations are less explored due to their corresponding challenges. Renormalization theory shows that irrelevant deformations require an infinite number of counterterms for the computation of physical quantities, resulting in Lagrangian densities, including an infinite series of terms, with significant ambiguities arising in the process. Nevertheless, recent advances show that these issues can be overcome for certain special classes of irrelevant deformations, making the problem more manageable. This class of deformations is known as $\ttb$ deformations \cite{Cavaglia:2016oda, Smirnov:2016lqw}, which can be defined in terms of the Lagrangian flow equation
\begin{equation}
    \frac{\partial \mathcal{L}^{(\tau)}}{\partial \tau} = -\frac{1}{\pi^2} \ttb\,,
\end{equation}
where $\mathcal{L}^{(\tau)}$ denotes the Lagrangian density of the theory, $\tau$ is a flow parameter, and the $\ttb$ operator is defined as
\begin{equation}
   \ttb =-\pi^2 \operatorname{det} T_{\mu \nu}=-\pi^2\left(T_{x x} T_{y y}-T_{x y}^2\right) = T \overline{T}-\Theta^2 \,,
\end{equation}
where $T_{\mu\nu}$ is the energy-momentum tensor. At the quantum level, denoting by $\ket{n}$ an eigenstate of both energy and momentum, such that
\begin{equation}
   \mathcal{H}\ket{n}= E_n\ket{n} \,,\quad \mathcal{P}\ket{n} = P_n \ket{n}\,,
\end{equation}
the following factorization property holds \cite{Zamolodchikov:2004ce}:
\begin{equation}
    \bra{n}{\ttb}\ket{n} =  \bra{n} T \ket{n}  \bra{n} \overline{T}\ket{n} - \bra{n}\Theta\ket{n}^2\,.
\end{equation}
Previous research \cite{Zamolodchikov:1991vx, Mussardo:1999aj, Dubovsky:2012wk, Caselle:2013dra} has emphasized the special role of the $\ttb$ operator in integrable perturbations of conformal field theories (CFTs), suggesting connections with the exact S-matrix theory and the TBA. Within the context of 2D integrable models, $\ttb$-deformed field theories retain an infinite number of conserved charges, allowing for the exact computation of scattering amplitudes and other observables. In particular, the scattering matrix $\mathbf{S}(\theta)$ is modified via the introduction of a CDD (Castillejo-Dalitz-Dyson) factor, such that
\begin{equation}\label{redf}
 \mathbf{S}^{(\tau)}(\theta)=\mathbf{S}^{(0)}(\theta) \times e^{\imath \tau m^2 \sinh \theta}\,, 
\end{equation}
where $m$ is a mass scale, and $\theta$ is the relative rapidity. From the TBA equations, it is possible to show that the $\ttb$-deformed energy spectrum on the cylinder follows the inviscid Burgers equation of hydrodynamics  \cite{Cavaglia:2016oda, Smirnov:2016lqw},
\begin{equation}\label{burgers_eq}
\frac{\partial E_n^{(\tau)}( R)}{\partial \tau} =  E_n^{(\tau)}( R)\frac{\partial E_n^{(\tau)}( R)}{\partial  R}+\frac{P_n^2(R)}{R}\,,
\end{equation}
where $P_n=2\pi n/R$ is the (quantized) total momentum. Notice, however, that in a field theory confined to a finite cylinder geometry with Neumann boundary conditions,  only the states with $P_n(R) = 0$ contribute to the partition function. In the zero-momentum sector, equation \eqref{burgers_eq} can be implicitly solved by the following equation:
\begin{equation}\label{imp} E_n^{(\tau)}(R) = E_n^{(0)}(\tilde{R})\,,\quad \text{with} \quad \tilde{R} = R + \tau E_n^{(\tau)}\,. \end{equation}
Equation \eqref{imp} establishes an explicit connection between the deformation parameter $\tau$ and an auxiliary deformed geometry through the dynamical map $R \mapsto \tilde{R}$.

When the un-deformed theory is a conformal field theory (CFT), a general solution to the Burgers equation \eqref{burgers_eq} for zero-momentum states is provided by
\begin{equation}\label{E:cft}
    E_n^{(\tau)}(R) = -\frac{R}{2\tau}\pm \sqrt{\left(\frac{R}{2\tau}\right)^2 -\frac{2\pi}{\tau}{\left(\frac{c_{\mathrm{eff}}}{12}-\Delta\right)}}\,,
\end{equation}
where $c_{\mathrm{eff}}$ denotes the effective central charge, and $\Delta$ is the scaling dimension of the corresponding CFT operator on the plane. The physical states of the deformed CFT correspond to those with the positive sign in \eqref{E:cft}. The non-physical branch (negative sign) merges with the physical one at a branch point, which, for the ground state, corresponds to the well-known Hagedorn-type singularity appearing in string theory.
For $c_\text{eff} = D-2$, with $D=26$, the physical spectrum coincides with the spectrum of the Nambu-Goto string in a $26$-dimensional target space obtained through light-cone quantization.\footnote{It must be noted that the 
spectrum of the form $E_n^{(\tau)}(R) +R/2\tau$ is known to be incompatible with target space Poincaré symmetry, except for $D=26$, where the Virasoro algebra is free of anomalies, and possibly $D=3$ (see, for example, the related comments in  \cite{Aharony:2010db}, showing that deviations from the Nambu-Goto action can only occur at eight-derivative order — or higher — in three space-time dimensions, as well as \cite{Dubovsky:2015zey,Chen:2018keo}). However, $\ttb$ deformations of generic CFTs can be interpreted as non-critical string models \cite{Callebaut:2019omt}.
} A key subtlety arises concerning the sign of the deformation parameter $\tau$: different choices lead to significantly different physical behaviors.

\begin{figure}
\centering
\includegraphics[scale=.20]{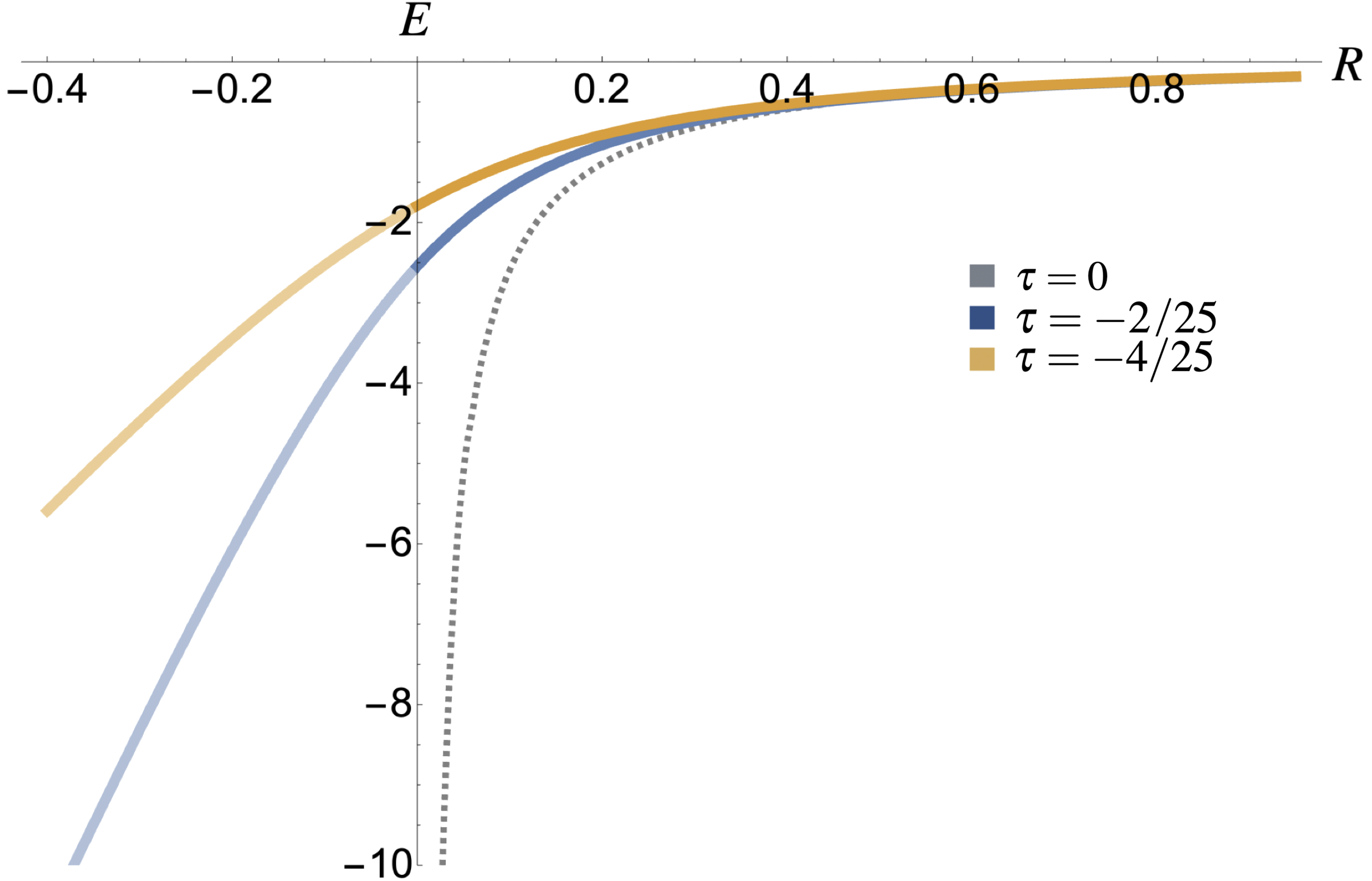}\\\quad\\\quad\\
\includegraphics[scale=.19]{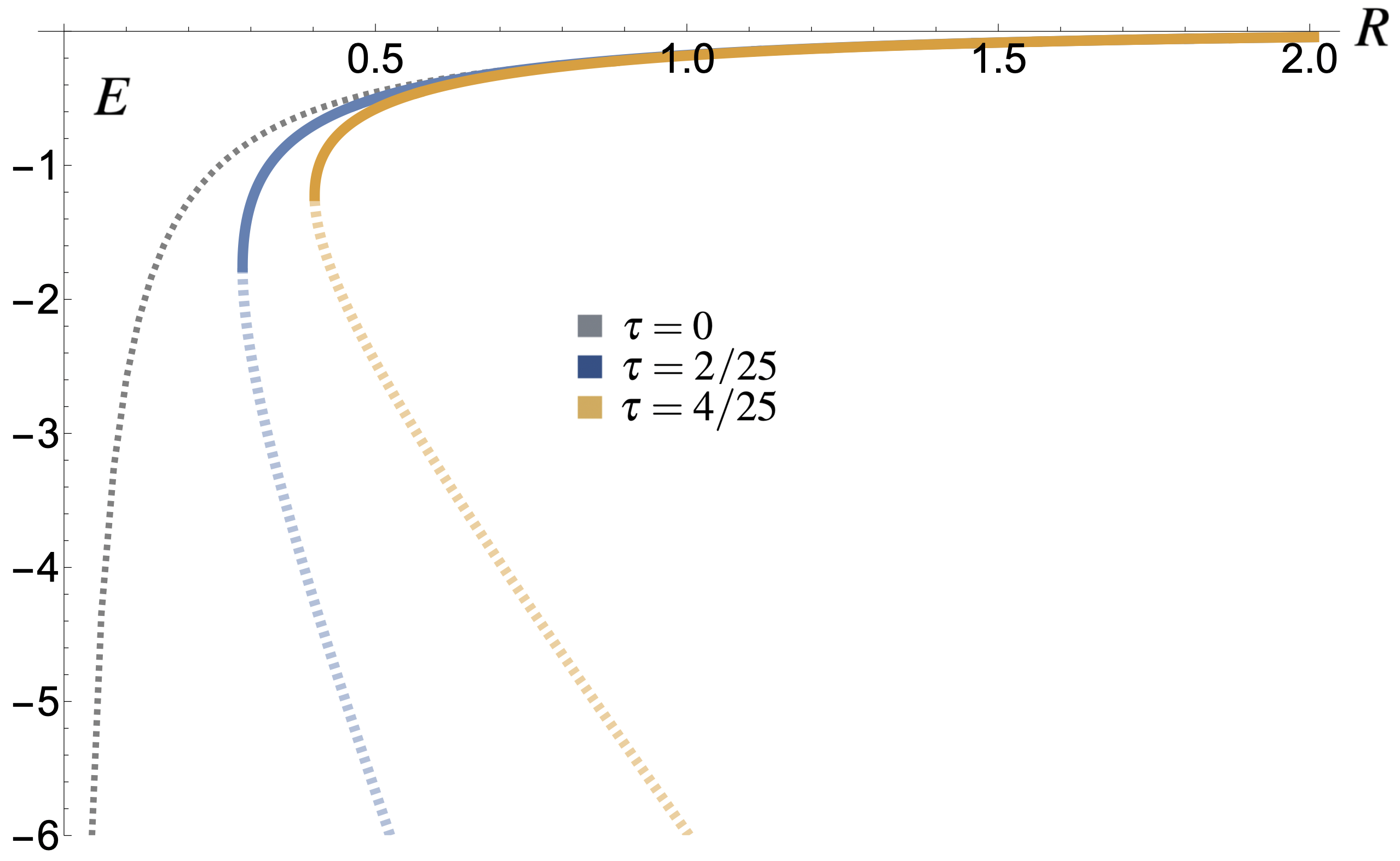}\\\quad\\
\caption{
The $\ttb$-deformed ground-state spectrum for the free fermion case is presented. In the first panel, the lighter-colored lines correspond to the forbidden region where $R<0$.
In the second panel, the dashed-colored lines represent the analytic continuation around the solution branch. This non-physical solution depicts the behavior of the ground-state spectrum when extended to the additional branch in the solution space.
} 
\label{spectrum}
\end{figure}

\begin{itemize}
    \item For negative $\tau$, an analysis of the density of states in the deep UV regime reveals behavior characteristic of a class of non-local theories known as little string theory, which is dual to gravitational theories on a linear dilaton background \cite{Giveon:2017nie}. The deformation effectively acts as a UV cutoff, softening high-energy divergences. However, this can be problematic from the perspective of causality.
    \item For positive $\tau$, the deformed energy levels can become complex at high energies, signaling a breakdown of unitarity or the presence of a Hagedorn-type transition (an upper limit to the temperature of a system due to an exponential growth in the number of states), arising from a pair of square-root branch points in the complex $R$-plane (see Figure \ref{spectrum}). This phenomenon is anticipated in theories of two-dimensional quantum gravity, where a minimum length scale exists \cite{McGough:2016lol, Guica:2019nzm}.
\end{itemize}

By exploiting modular invariance, the flow equations for the $\ttb$-deformed $g$-function can be derived \cite{Cardy:2018sdv, Jiang:2021jbg}: 
\begin{equation}\label{flow} \frac{\partial s_n^{(\tau)}(R)}{\partial \tau} = E_n^{(\tau)}(R)\frac{\partial s_n^{(\tau)}(R)}{\partial R} + R\frac{\partial}{\partial R}\frac{E_n^{(\tau)}(R)}{R}\,,\quad s_n^{(\tau)} = \ln \left(g_{n,\ket{w_1}}^{(\tau)}g_{n,\ket{w_2}}^{(\tau)}\right)\,. \end{equation}
The interpretation of this flow in the classical limit and its potential applications to the Nambu-Goto spectrum remain important areas for further investigation. Starting from a conformal field theory, the exact form of the $\ttb$-deformed boundary entropy $s_n^{(\tau)}$ is obtained by incorporating an extra factor which depends on the energy $E_n^{(\tau)}(R)$ \cite{Cavaglia:2016oda}:
\begin{equation}\label{result} s^{(\tau)}_{n,\text{CFT}}(R) = \ln\left(\braket{w_1|\psi_n}^\text{CFT}\right) +\ln\left(\braket{w_2|\psi_n}^\text{CFT}\right) +\ln\left(\frac{R}{R + 2\tau E_n^{(\tau)}(R)}\right)\,.
\end{equation}
The origin of the last term on the right-hand side of the equation in (\ref{result}) can be traced back to the additional contribution, compared to the initial proposal of \cite{LeClair:1995uf}, identified in \cite{Dorey:2004xk} (see also \cite{Pozsgay:2010tv, Kostov:2018dmi}), which is required for the equality between the closed and open channel representations of the cylinder partition function.
More details on the TBA-related approach to the $g$-function will be given in section \ref{exactg} below.

On the other hand, one can employ the \emph{method of characteristics} to derive the \emph{general solution} to the flow equation \eqref{flow} \cite{Jiang:2021jbg} 
\begin{equation}\label{jung}
s_n^{(\tau)}(R)=s_n^{(0)}\left(R+\tau E_n^{(\tau)}(R)\right)+\ln{\left(\frac{R+\tau R \partial_R E_n^{(\tau)}(R)}{R+\tau E_n^{(\tau)}(R)}\right)}\,,
\end{equation}
where the first term on the right-hand side represents the un-deformed entropy on the deformed geometry $\tilde{R} = R+\tau E_n^{(\tau)}(R)$. 
The expressions \eqref{jung} and \eqref{result} coincide when the Nambu-Goto type spectrum \eqref{E:cft} is involved.

Beginning from the observation that, within the TBA framework, the deformation is equivalent to the modification of the S-matrix through a CDD factor \eqref{redf}, the objective of this note is to derive the expression \eqref{jung} exclusively by employing the TBA definition of the exact $g$-function for a generic massive theory (see Section \ref{exactg}). 
While the calculations can be performed analytically in general, in this brief review, we focus on the specific case of the Ising model (massive Majorana fermion).

\section{The Exact $g$-function}
\label{exactg}
In this section, to simplify the discussion and avoid unnecessary complications, we will restrict our analysis to a purely elastic scattering theory involving a single particle species of mass $m$. The discussion can be easily extended to more general scattering theories and configurations \cite{Dorey:1997yg,Dorey:1999cj, Bajnok:2001ug, Corrigan:2000fm,Ahn:2003st,  Dorey:2005ak,Ahn:2005fj,Kormos:2007qb}.

It was first proposed by \cite{LeClair:1995uf}, that the boundary entropy should be given by an expression of the form:
\begin{equation}\label{g:inc}
\int_\mathbb{R}\frac{\mathrm{d}\theta}{4\pi}\Theta(\theta)\ln\left(1+e^{-\epsilon(\theta|R)}\right)\,,
\end{equation}
where $\epsilon(\theta|R)$, satisfying a non-linear integral equation from the TBA, is given by
\begin{equation}\label{TBA}
\epsilon(\theta|R)=mR\cosh\theta-\frac{1}{2\pi}\phi\ast\ln\left(1+\text{e}^{-\epsilon}\right)(\theta)\,,
\end{equation}
where we introduced the compact notation
\begin{equation}
    f\ast  g (\theta) =\int_\mathbb{R} \mathrm{d}\theta' f(\theta-\theta') g(\theta')\,.
\end{equation}
The function $\Theta(\theta)$ is defined as
\begin{equation}
\Theta(\theta)=-\imath\frac{\mathrm{d}}{\mathrm{d}\theta}\ln(\mathbf{R}_{\ket{w_1}}(\theta)\mathbf{R}_{\ket{w_2}}(\theta))+\imath\frac{\mathrm{d}}{\mathrm{d}\theta}\ln\,\mathbf{S}(2\theta)-2\pi\delta(\theta)\,,
\end{equation}
where $\mathbf{S}(\theta)$ and $\mathbf{R}_{\ket{w}}(\theta)$ denote the scattering and reflection matrices, respectively. The kernel $\phi(\theta)$ is directly related to the scattering matrix through
\begin{equation}
\phi(\theta)=-\imath\frac{\mathrm{d}}{\mathrm{d}\theta}\ln\,\mathbf{S}(\theta)\,.
\end{equation} 
However, it was later observed with a careful numerical analysis in \cite{Dorey:1999cj} that the expression \eqref{g:inc} is incomplete due to the exclusion of a boundary-independent term. By employing an $n$-particle cluster expansion, reminiscent of form-factor techniques for correlation functions, the missing term was finally recovered in \cite{Dorey:2004xk}. The complete expression for the boundary entropy can be written in the form
\begin{equation}\label{bulk}
s^{(0)}(R)=\int_\mathbb{R}\frac{\mathrm{d}\theta}{4\pi}\Theta(\theta)\ln\left(1+e^{-\epsilon(\theta|R)}\right)+\sum_{k>0}\mathbf{I}_k(R)\,,
\end{equation}
where
\begin{align}\label{first}
&\mathbf{I}_1(R)=\int_\mathbb{R}\frac{\mathrm{d}\theta}{2\pi}\frac{\phi(2\theta)}{1+e^{\epsilon(\theta|R)}}\,,\\
&\mathbf{I}_{k>1}(R)=\frac{1}{k}\left(\prod_{p=1}^k\int_\mathbb{R}\frac{\mathrm{d}\theta_p}{2\pi}\right)\frac{\phi(\theta_1+\theta_2)}{1+e^{\epsilon(\theta_1|R)}}\left(\prod_{q=2}^k\frac{\phi(\theta_q-\theta_{q+1})}{1+e^{\epsilon(\theta_q|R)}}\right)\delta_{\theta_{k+1},\theta_1}\,.
\end{align}
We also recall that the ground-state energy of the theory on a circle of circumference $R$ is given in terms of the pseudo-energy by \cite{Zamolodchikov:1989cf}
\begin{equation}
E_0^{(0)}(R)=-m\int_{\mathbb{R}}\frac{\mathrm{d}\theta}{2\pi}\cosh\theta\ln\left(1+e^{-\epsilon(\theta|R)}\right)\,.
\end{equation}

\section{Deforming the Entropy}

We can implement the deformation by modifying the scattering properties of the theory \cite{Dubovsky:2013ira, Caselle:2013dra, Cavaglia:2016oda, Smirnov:2016lqw},
\begin{equation} 
\begin{aligned}
&\mathbf{S}\,\longrightarrow\,\mathbf{S}^{(\tau)}\,,\\
&\mathbf{R}\,\longrightarrow\,\mathbf{R}^{(\tau)}\,.
\end{aligned}
\end{equation}
In this framework, the infinite sum of bulk contributions is always deformed, even in cases such as the free fermion, where $\mathbf{S}^{(0)}=-1$, and the contribution was originally absent.

As discussed in the introduction, the $\ttb$-deformed scattering matrix is purely a phase, with no poles or zeros,
\begin{equation}\label{matrix}
\mathbf{S}^{(\tau)}(\theta) = \mathbf{S}^{(0)}(\theta) \times e^{\imath\tau m^2 \sinh\theta}\,,
\end{equation}
so that
\begin{equation}\label{kernel}
\phi^{(\tau)}(\theta) = \phi^{(0)}(\theta) + \tau m^2 \cosh\theta\,.
\end{equation}
By using this expression for the kernel in the TBA equation \eqref{TBA}, and after straightforward manipulations, we obtain \cite{Cavaglia:2016oda}:
\begin{equation}\label{epsilon}
\epsilon^{(\tau)}(\theta|R) = \epsilon^{(0)}(\theta|\tilde{R})\,,
\end{equation}
where $\tilde{R} = R + \tau E_0^{(\tau)}$. This general property is central to the remarkable result \eqref{imp}, allowing us to derive the deformation by explicitly modifying the scale $R$ in accordance with \eqref{epsilon}.

The bulk corrections are modified in a non-trivial manner, as they depend on the pseudo-energy and the kernel itself.
To clarify this concept, let us consider the first contribution \eqref{first} for simplicity. Explicitly, one has
\begin{equation}
\mathbf{I}_1^{(0)}(R)=\int_\mathbb{R}\frac{\mathrm{d}\theta}{2\pi}\frac{\phi^{(0)}(2\theta)}{1+e^{\epsilon(\theta|R)}}\,\longrightarrow\,\mathbf{I}_1^{(\tau)}(R)=\int_\mathbb{R}\frac{\mathrm{d}\theta}{2\pi}\frac{\phi^{(\tau)}(2\theta)}{1+e^{\epsilon^{(\tau)}(\theta|R)}}\,,
\end{equation}
where, using definition \eqref{kernel} and property \eqref{epsilon},
\begin{align}
{\bf I}^{(\tau)}_1(R)=&\int_\mathbb{R}\frac{\mathrm{d}\theta}{2\pi}\frac{\phi^{(0)}(2\theta)}{1+e^{\epsilon^{(0)}(\theta|\tilde{R})}}+\tau m^2\int_\mathbb{R}\frac{\mathrm{d}\theta}{2\pi}\frac{\cosh{2\theta}}{1+e^{\epsilon^{(\tau)}(\theta|R)}}\\
=&\,\,{\bf I}_1^{(0)}(\tilde{R})+\fnc_1^{(\tau)}(R)\,.
\end{align}
In particular,
\begin{equation}
\fnc_1^{(\tau)}(R)\equiv\tau m^2\int_\mathbb{R}\frac{\mathrm{d}\theta}{2\pi}\frac{\cosh{2\theta}}{1+e^{\epsilon^{(\tau)}(\theta|R)}}=\tau m^2\int_\mathbb{R}\frac{\mathrm{d}\theta}{2\pi}\frac{\cosh{2\theta}}{1+e^{\epsilon^{(0)}(\theta|\tilde{R})}}\,,
\end{equation}
where we have used once again the property \eqref{epsilon} in the last identification. Similarly, we can define $\forall k>0$, functions $\fnc_k^{(\tau)}$ encoding the whole explicit $\tau$-dependence. Compactly, one can write
\begin{equation}
{\bf I}_k^{(0)}(R)\,\longrightarrow\,{\bf I}^{(\tau)}_k(R)={\bf I}^{(0)}_k(\tilde{R})+\fnc_k^{(\tau)}(R)\,.
\end{equation}
Meanwhile, the general form of the boundary-perturbing reflection factor is given by \cite{Caselle:2013dra, Cavaglia:2016oda}:
\begin{equation}\label{R:tt}
\mathbf{R}^{(\tau)}(\theta) = \mathbf{R}^{(0)}(\theta) \times e^{\imath\tau m^2 \frac{\sinh(2\theta)}{2}} \prod_{a \ge 1} e^{\delta_a (i m \sinh\theta)^{2a-1}}\,.
\end{equation}
Each factor in the product on the right-hand side of \eqref{R:tt} corresponds to a specific boundary perturbation with coupling $\delta_n$ \cite{Cavaglia:2016oda}. Combining these expressions, we obtain:
\begin{equation}\label{theta}
\Theta(\theta) = \Theta^{(0)}(\theta) + 2 \sum_{a \ge 1} (2n-1)(-1)^{n+1} \delta_a m^{2a-1} \cosh\theta (\sinh\theta)^{2a-2}\,.
\end{equation}
Interestingly, the function $\Theta(\theta)$ is explicitly independent of $\tau$, due to a cancellation mechanism between the scattering factor and the fundamental boundary reflection factor. Using this property, we can express the deformed entropy as
\begin{equation}
\begin{aligned}
&\int_\mathbb{R} \frac{\mathrm{d}\theta}{4\pi} \Theta(\theta) \ln\left(1 + e^{-\epsilon^{(\tau)}(\theta|R)}\right) + \sum_{k>0} {\bf I}_k^{(\tau)}(R) + \\
&\int_{\mathbb{R}} \frac{\mathrm{d}\theta}{2\pi} \sum_{a \ge 1} (2a-1)(-1)^{a+1} \delta_a m^{2a-1} \cosh\theta (\sinh\theta)^{2a-2} \ln\left(1 + e^{-\epsilon^{(\tau)}(\theta|R)}\right)\,.
\end{aligned}
\end{equation}
When setting $\delta_a = 0$, we can apply the property \eqref{epsilon} to express
\begin{equation}
s^{(\tau)}(R) = s^{(0)}(\tilde{R}) + \sum_{k>0}\fnc_k^{(\tau)}(R)\,.
\end{equation}
Even in this relatively simple setup, where only a single particle species is involved, it is far from obvious how the infinite corrections in $\tau$ can be reduced to the expected result \eqref{jung}, i.e.
\begin{equation}
\sum_{k>0}\fnc_k^{(\tau)}(R)=\ln{\left(\frac{R+\tau R\partial_R E_0^{(\tau)}(R)}{R+\tau E_0^{(\tau)}(R)}\right)}\,.
\end{equation}
This holds in the massless limit, as demonstrated in \cite{Cavaglia:2016oda}, but becomes highly non-trivial in the general case.

\section{Deforming the Free Fermion}

The aim of this section is to present a concrete example of how the exact expression \eqref{jung} can be systematically reconstructed within the context of the free fermion case.

The challenge arises from the infinite sum of bulk corrections that contribute to the exact entropy as a result of the deformation. Using the kernel \eqref{kernel}, where 
\begin{equation}
\phi_\text{Fermion}^{(0)} \equiv 0\,,
\end{equation}
we can express these corrections (${\bf I}^{(0)}_k \equiv 0$) as:
\begin{align}
&\fnc_1^{(\tau)}=\tau m^2\int_{\mathbb{R}}\frac{\mathrm{d}\theta}{2\pi}\frac{1}{1+e^{\tilde{r}\cosh\theta}}\cosh(2\theta)\,,\\
&\fnc_{k>1}^{(\tau)}=\frac{(\tau m^2)^k}{k}\left(\prod_{p=1}^k\int_{\mathbb{R}}\frac{\mathrm{d}\theta_p}{2\pi}\right)\frac{\cosh(\theta_1+\theta_2)}{1+e^{\tilde{r}\cosh\theta_1}}\left(\prod_{q=2}^k\frac{\cosh(\theta_q-\theta_{q+1})}{1+e^{\tilde{r}\cosh\theta_q}}\right)\delta_{\theta_{k+1},\theta_1}\,,
\end{align}
where we have used the property \eqref{epsilon} to represent the deformed pseudo-energy as $\epsilon^{(\tau)}(\theta)\left|_{\text{Fermion}}\right.=\tilde{r}(\tau)\cosh\theta$, with $\tilde{r}/m\equiv \tilde{R}$.

Remarkably, much like in the case of CFTs \cite{Cavaglia:2016oda}, we are able to \emph{resum} these contributions. Through explicit calculations involving trigonometric functions, we have derived a fundamental identity:
\begin{equation}\label{sum}
\sum_{k>0}\fnc_k^{(\tau)} =\sum_{n\ge0}\mathbf{M}_n^{(\tau)}\,,
\end{equation}
where 
\begin{align}
\mathbf{M}_n^{(\tau)}=&\frac{2^{-2 n-3} \tau ^{2 n+1}}{(n+1) (2 n+1)} \left\{\left(V+U\right)^{2 n+1} [(2 n+1) \tau  (V+U)+4 (n+1)]\right.\notag\\
&\left.-\left(V-U\right)^{2 n+1} [(2 n+1) \tau  (V-U)+4 (n+1)]\right\}\,,
\end{align}
and 
\begin{equation}
U=m^2\int_{\mathbb{R}}\frac{\mathrm{d}\theta}{2\pi}\frac{\cosh 2\theta}{e^{\tilde{r}\cosh\theta}+1}\,,\quad V=m^2\int_{\mathbb{R}}\frac{\mathrm{d}\theta}{2\pi}\frac{1}{e^{\tilde{r}\cosh\theta}+1}\,.
\end{equation}
This identity can be straightforwardly verified by applying Werner’s second formula,
\begin{equation}
   \cosh (x+y) \cosh (x-y)=\frac{1}{2}(\cosh 2 x+\cosh 2 y) \,, 
\end{equation}
and using the following relation:
\begin{equation} \left(\prod_i\int_{-c}^{c} dx_i\right)\cosh{\sum_i x_i} = \left(\prod_i\int_{-c}^{c} dx_i\right)\prod_i \cosh x_i\,. \end{equation}
The sum in \eqref{sum} is well-defined, and we get 
\begin{equation}\label{iden1}
    \begin{split}
    \sum_{k > 0} \fnc_k^{(\tau)}(R) &=  \frac{1}{2}\ln\left[\left(\frac{(U-V)\tau+2}{(U+V)\tau-2}\right)^2\right]=\frac{1}{2}\ln{\left[\left(\frac{R+\tau R \partial_R E_0^{(\tau)}(R)}{R+\tau E_0^{(\tau)}(R)}\right)^2\right]}\,.    
    \end{split}
\end{equation}
The final identification in \eqref{iden1} follows from straightforward manipulations of the total energy for the deformed fermion, i.e.,
\begin{align}
E_0^{(\tau)}&=-m\int_{\mathbb{R}}\frac{\mathrm{d}\theta}{2\pi}\cosh\theta\ln\left(1+e^{-\tilde{r}\cosh(\theta)}\right)\notag\\
&=-m^2R\int_{\mathbb{R}}\frac{\mathrm{d}\theta}{2\pi}\sinh^2\theta\frac{1}{e^{\tilde{r}\cosh\theta}+1}-m^2\tau E_0^{(\tau)}\int_{\mathbb{R}}\frac{d\theta}{2\pi}\sinh^2\theta\frac{1}{e^{\tilde{r}\cosh\theta}+1}
\end{align}
and
\begin{align}
\frac{\partial}{\partial R}E_0^{(\tau)}&=m^2\int_{\mathbb{R}}\frac{\mathrm{d}\theta}{2\pi}\cosh^2\theta\frac{1}{e^{\tilde{r}\cosh\theta}+1}\notag\\
&+m^2\tau\frac{\partial}{\partial R}E_0^{(\tau)} \int_{\mathbb{R}}\frac{\mathrm{d}\theta}{2\pi}\cosh^2\theta\frac{1}{e^{\tilde{r}\cosh\theta}+1}\,.
\end{align}
In particular, we can express
\begin{equation}
-\frac{E_0^{(\tau)}}{R}=\frac{u}{1+\tau u}\,,\quad\frac{\partial}{\partial R}E_0^{(\tau)}=\frac{v}{1-\tau v}\,,
\end{equation}
with
\begin{equation}
U = u + v\,,\quad V = u - v\,.
\end{equation}
% The result \eqref{iden1} matches the exact expression \eqref{jung} across the entire complex plane of
% \begin{equation}\notag
% \frac{R + \tau R\partial_R E_0^{(\tau)}(R)}{R + \tau E_0^{(\tau)}(R)}\,,\quad\forall \tau\,.
% \end{equation}
By assembling all the contributions, one precisely recovers the result \eqref{jung} for the ground state. The explicit calculation provides substantial control over each contribution. The generalization to excited states is obtained by replacing the integration contour $\mathbb{R}$ with a contour encircling zeros of $1+e^{-\epsilon(\theta)}$ \cite{Dorey:1997rb}, this means $E_0^{(\tau)}\rightarrow E_n^{(\tau)}$. It has thus been shown that the solution obtained via the method of characteristics is entirely consistent with the one derived using the more traditional TBA-related approach.

\section{Discussion \& Numerics}
\begin{figure}[th!]
\centering
\includegraphics[scale=.20]{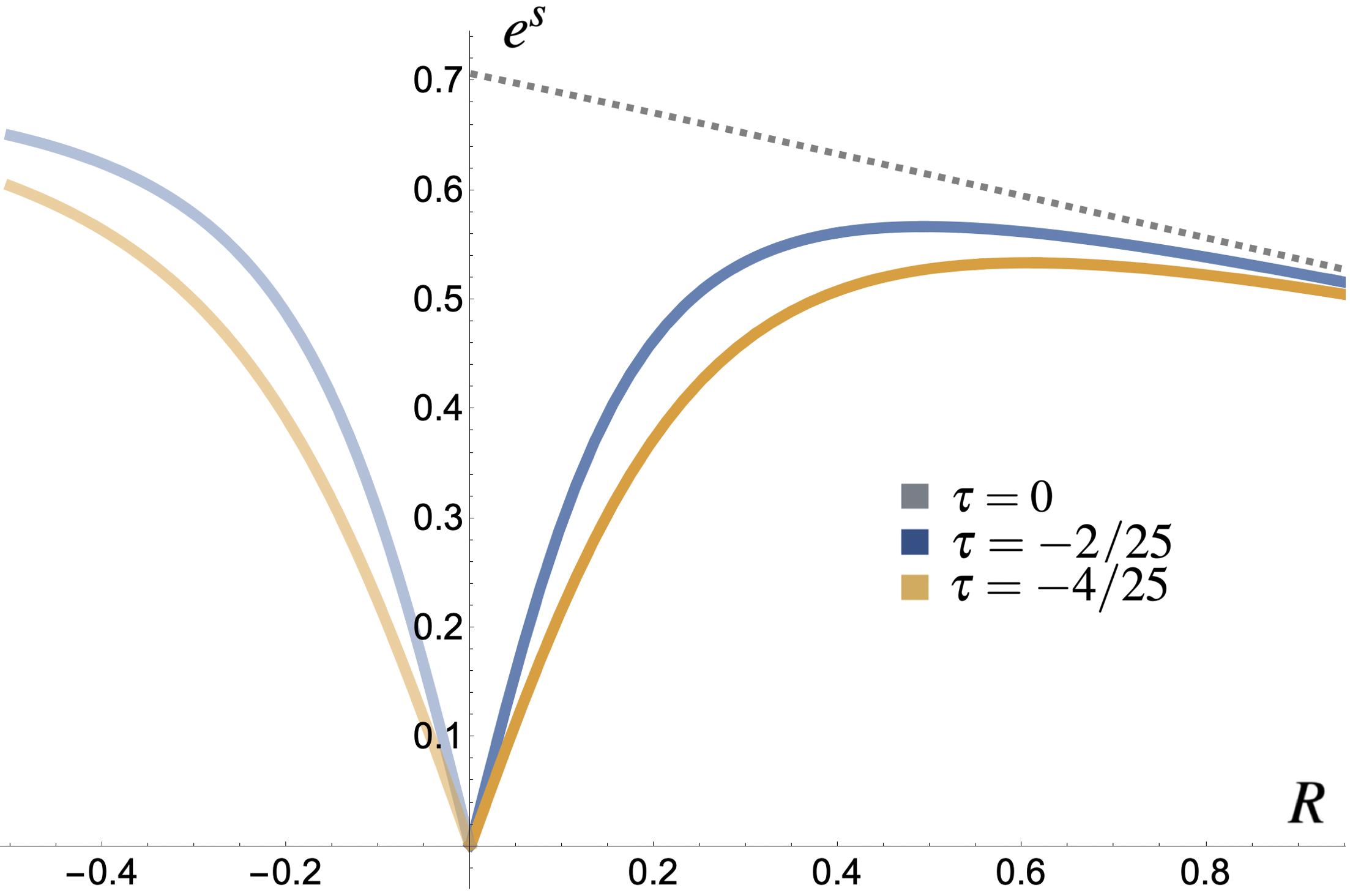}\\\quad\\
\caption{The $\ttb$-deformed ground state entropy in the Majorana free fermion model, under ``fixed-free'' boundary conditions \cite{Ghoshal:1993tm}, is examined here with $\delta_a = 0$ for simplicity. In the panel, the lighter-colored lines indicate the forbidden region corresponding to $R<0$.} 
\label{entropy1}
\end{figure}

\begin{figure}[th!]
\centering
\includegraphics[scale=.20]{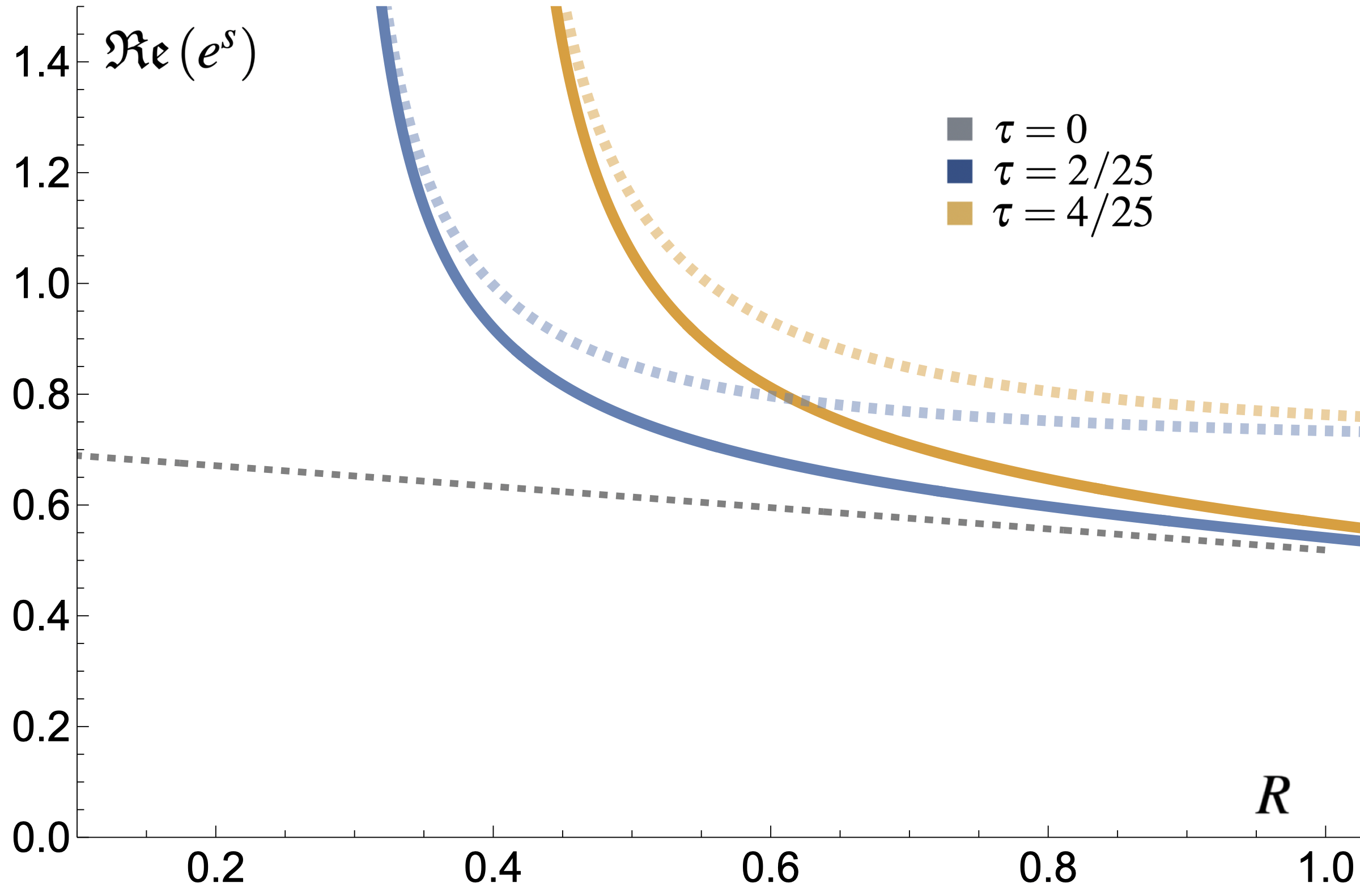}\\\quad\\
\caption{
In the panel, the deformed ground state entropy in the free fermion model, under ``fixed-free'' boundary conditions, is examined ($\delta_a = 0$). The analytic continuation around the solution branch is highlighted by dashed-colored lines.}   
\label{entropy2}
\end{figure}
\begin{figure}[th!]
\centering
\includegraphics[scale=.20]{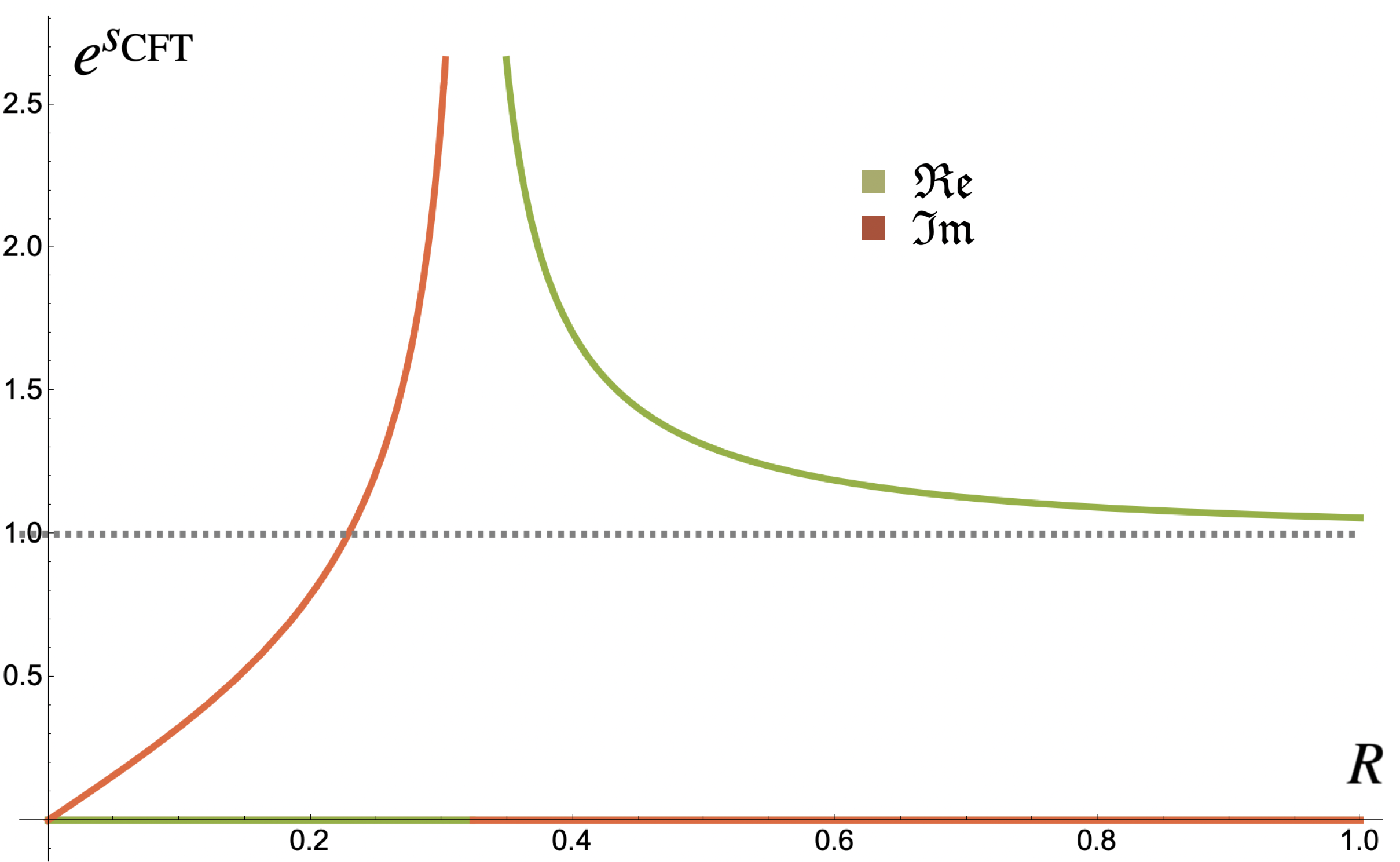}\\\quad\\
\caption{The $\ttb$-deformed ground state entropy in the CFT model, under ``free'' boundary conditions \cite{Dorey:2009vg}, is examined here at $\tau=1/10$. The un-deformed value is reported in dashed-gray}   
\label{cft}
\end{figure}

In this note, we investigate the effects of $\ttb$ deformation on a simple model with boundaries. Our methodology can be easily generalized. The \emph{resummation procedure} employed here may also be extended to incorporate more general scattering matrices than that specified in \eqref{matrix}. Let us notice that the flow \eqref{flow} equation can also be proven true in the case $\delta_a\ne0$.

A numerical evaluation of the flow equations reveals striking similarities between the deformed spectrum and the entropy. For the sake of simplicity, we consider ``fixed-free'' boundary conditions in the un-deformed theory \cite{Ghoshal:1993tm},
\begin{align}\label{boundary}
&\mathbf{R}_{\ket{w_1}}(\theta)=\mathbf{R}_{\ket{\text{fixed}}}(\theta)=\imath\tan(\imath\pi/4-\theta/2)\,,\\
&\mathbf{R}_{\ket{w_2}}(\theta)=\mathbf{R}_{\ket{\text{free}}}(\theta)=-\imath\,\text{cotan}(\imath\pi/4-\theta/2)
\end{align}
which simplifies to
\begin{equation}
\Theta(\theta)=-2\pi\delta(\theta)\,,
\end{equation}
assuming then $\delta_a=0$ in \eqref{theta}.\footnote{We work in natural units where $m=1$.}

For $\tau < 0$, the ground state function $\exp s_0$ exhibits no singularities, but deviates from its behavior at $\tau = 0$, vanishing linearly in the ultraviolet (UV) regime as $R \rightarrow 0$ (see Figure \ref{entropy1}). This behavior, typically manifesting at macroscopic scales, can be expressed as:
\begin{equation}\label{behav} \lim_{R\rightarrow0} \frac{\braket{w_1|\psi_0}^{(\tau<0)}}{\sqrt{\braket{\psi_0|\psi_0}^{(\tau<0)}}} = 0\,. \end{equation}
This result is straightforwardly recovered in the massless limit, where the relevant expressions are explicit.

For $\tau > 0$, $\exp s_0^{(\tau)}$ reveals the existence of a minimum length scale $R_c$, below which such quantity becomes complex. In Figures \ref{entropy2} and \ref{cft}, this corresponds to a singularity of the form $(R - R_c)^{-1/2}$. To probe the behavior for $R < R_c$, we can consider the CFT case, where $\exp s_{0,\text{CFT}}^{(\tau)}$ becomes purely imaginary, and again tends to zero linearly as $R \rightarrow 0$. For $R > R_c$, in addition to the physical solution, $\exp s_0^{(\tau)}$ has a second real branch solution (depicted by the dashed-colored lines in Figure \ref{entropy2}). This non-physical branch arises from the analytic continuation of the spectrum around the branch point (see Figure \ref{spectrum}). In the massless limit, where $R_c = \sqrt{2\pi \tau c_\text{eff}/3}$, the two branches (for $R > R_c$) collapse into a single one.

In general, unlike the $\tau = 0$ cases, the $\ttb$-deformed quantities $\exp s_{n,\text{CFT}}^{(\tau)}$ vanish linearly as $R \rightarrow 0$, suggesting the need for an appropriate renormalization of the partition function. Moreover, when transitioning from the ground state to the excited states, the qualitative behavior for $\tau < 0$ and $\tau > 0$ is exchanged.

A comprehensive numerical investigation could provide further insights into these aspects (see, for example, \cite{Camilo:2021gro}).

\section{Open Problems and Future Directions}
The effect of the $\ttb$ deformation on boundary conditions, particularly on the boundary term of the classical action, as well as the nature and properties of the associated deforming operators, remains a critical open question in this research field (see, e.g., \cite{Babaro:2018cmq, Jiang:2021jbg}). A thorough investigation of these aspects is a crucial first step toward understanding the implications of the broader class of $\ttb$-like deformations \cite{Smirnov:2016lqw, Bzowski:2018pcy, Guica:2019vnb, Conti:2019dxg,LeFloch:2019wlf}, including the more recently introduced classically marginal root-$\ttb$ deformation \cite{Conti:2022egv, Ferko:2022cix, Babaei-Aghbolagh:2022leo, Borsato:2022tmu, Ebert:2023tih, Babaei-Aghbolagh:2024hti, Tsolakidis:2024wut}, in systems with boundaries. The geometric interpretation of these more general deformations is particularly intriguing, as it may uncover further connections to topological gravity and random geometry \cite{Dubovsky:2018bmo, Cardy:2018sdv, Hirano:2020nwq}.

An important open question worth highlighting concerns the so-called ODE/IM correspondence (see \cite{Dorey:1998pt, Bazhanov:1998wj, Dorey:2007zx, Lukyanov:2010rn, Dorey:2019ngq}). This framework establishes a precise dictionary between classical and quantum quantities in 2D integrable field theories. As shown in \cite{Aramini:2022wbn}, this quantum/classical correspondence is preserved under $\ttb$ deformation. However, the extension of the ODE/IM correspondence to the case of boundaries and crosscaps \cite{Caetano:2020dyp,Caetano:2021dbh} in the presence of $\ttb$-like deformations has not yet been explored and remains an intriguing open problem.

Notably, boundary problems—together with the simple Ising model example discussed in this note—naturally link with recent advances in the study of non-invertible symmetries, a topic that has garnered considerable interest in the scientific community (see \cite{Shao:2023gho} for a comprehensive review). See also \cite{Yan:2024yrw} for a discussion related to non-invertible symmetries emerging in deformed Ising models on the lattice, which share remarkable analogies with $\ttb$ flows in the continuous limit. In particular, the analytical tools offered by the $\ttb$ formalism may provide valuable insights into the space of topological defects, which can be understood as the insertion of boundary conditions.

Finally, extending the current framework to field theories in higher spacetime dimensions represents a promising direction for future research (see, for instance, \cite{Taylor:2018xcy, Conti:2022egv, Morone:2024ffm,Blair:2024aqz}). 
\bigskip

\textbf{Acknowledgements} -- 
We thank Changrim Ahn, Zoltan Bajnok, Patrick Dorey, Clare Dunning, Paul Pearce, Francesco Ravanini, Marton Kormos, and Gabor Takacs for insightful discussions. This work received partial support from the INFN project SFT and the PRIN Project No. 2022ABPBEY, ``Understanding quantum field theory through its deformations''.

Roberto Tateo thanks the organizers of the MATRIX Research Program ``Mathematics and Physics of Integrability (MPI2024)'' for the invitation and its participants for the stimulating atmosphere and discussions. He also expresses his gratitude to the Mathematical Research Institute MATRIX and the Australian Mathematical Sciences Institute for their financial support during his visit to Australia.

\bibliography{references}
\end{document}